\documentclass[aps,prd,twocolumn,amsmath,amssymb,amsfonts,nofootinbib,superscriptaddress,altaffilletter]{revtex4-1}
\usepackage{array}
\usepackage{graphicx}
\usepackage{cellspace}
\usepackage{amssymb}
\usepackage{amsmath}
\usepackage{longtable}
\usepackage{tabu}
\usepackage{ulem}
\usepackage{xcolor}
\usepackage{etoolbox}
\usepackage{supertabular}
\usepackage{dcolumn}
\usepackage{ wasysym }

\newcommand{\Freq}{f}
\newcommand{\fdot}{{\dot{\Freq}}}

\newtoggle{fullauthorlist}
\toggletrue{fullauthorlist}
\newcommand{\RNum}[1]{\uppercase\expandafter{\romannumeral #1\relax}}

\newtoggle{endauthorlist}
\toggletrue{endauthorlist}
\usepackage[utf8]{inputenc}
\DeclareUnicodeCharacter{2009}{-}
\DeclareUnicodeCharacter{2212}{-}

\begin{document}

\title{ 
Deep learning for clustering of continuous gravitational wave candidates \RNum{2}: identification of low-SNR candidates
}


\author{B. Beheshtipour}
\email{b.beheshtipour@aei.mpg.de}
\affiliation{Max Planck Institute for Gravitational Physics (Albert Einstein Institute), Callinstrasse 38, 30167 Hannover, Germany}
\affiliation{Leibniz Universit\"at Hannover, D-30167 Hannover, Germany}

\author{M.A. Papa}
\email{maria.alessandra.papa@aei.mpg.de}
\affiliation{Max Planck Institute for Gravitational Physics (Albert Einstein Institute), Callinstrasse 38, 30167 Hannover, Germany}
\affiliation{Leibniz Universit\"at Hannover, D-30167 Hannover, Germany}
\affiliation{University of Wisconsin Milwaukee, 3135 N Maryland Ave, Milwaukee, WI 53211, USA}

\begin{abstract}

Broad searches for continuous gravitational wave signals rely on hierarchies of follow-up stages for candidates above a given significance threshold. An important step to simplify these follow-ups and reduce the computational cost is to bundle together in a single follow-up nearby candidates. This step is called clustering and we investigate carrying it out with a deep learning network. In our first paper \cite{beh20}, we implemented a deep learning clustering network capable of correctly identifying clusters due to large signals. In this paper, a network is implemented that can detect clusters due to much fainter signals. These two networks are complementary and we show that a cascade of the two networks achieves an excellent detection efficiency across a wide range of signal strengths, with a false alarm rate comparable/lower than that of methods currently in use.  

\end{abstract}
\maketitle

\section{Introduction}
\label{intro}
The detection of gravitational waves from binary mergers  represents a historical breakthrough and marks the beginning of the era of gravitational wave astronomy \cite{abb16,nit19,lig18}. Continuous gravitational waves are ever-lasting nearly monochromatic gravitational waves which have not yet been detected, most probably  due to their weakness \cite{Lasky:2015uia,Authors:2019ztc,Pisarski:2019vxw,Dergachev:2019oyu,Dergachev:2019wqa}. They are expected from rotating compact objects when their shape or motion deviates from perfect axisymmetry \cite{Haskell:2015iia,Singh_2017,walin_1969,Abney_1996}. 

Broad-frequency surveys for continuous gravitational waves are routinely carried out \cite{Ming:2019xse,Papa:2020vfz,Steltner:2020hfd}, which investigate a very large number -- O(10$^{17}$) -- of possible waveforms, and require a so-called ``clustering" of the initial results. 

Why is clustering important? A gravitational wave signal or a disturbance may trigger not only one template but a number of nearby templates, to rise above the average noise level. This produces many ``candidates", each needing to be followed up and resulting in a high computing cost. To keep the computing cost in check one could increase the threshold that the detection statistic of a candidate has to exceed in order to be worthy of follow-up. But this obviously results in a loss of sensitivity. 
To tip the balance towards higher sensitivities at the same computing cost, clustering is used. 

Clustering identifies nearby candidates due to the same root-cause and bundles them together as one. Only the most significant candidate of the set  is followed-up and this reduces the computational cost. 

Clustering algorithms used in broad continuous waves surveys have evolved over time \cite{Steltner:2020hfd,ben15,pap16,sin17}. The shape of signal-clusters depends on the signal parameters and on the search parameters, and it is impossible to predict. The reason for this is that the mismatch reduction function can be computed analytically -- using the metric approach \cite{Prix:2006wm,Wette:2013wza,Wette:2015lfa} -- only at distances from the signal parameters that are too small to be informative for clustering, i.e. at distances smaller than the typical template-grid spacings and certainly much smaller than typical signal-cluster sizes. This means that in all deterministic clustering approaches, one has to resort to extensive Monte Carlos in order to see what signal clusters look like and then tune the clustering parameters accordingly. These studies have to be repeated for every new search and across the entire the parameter space searched, and are very time consuming.

In our first paper \cite{beh20} we explored the idea of using the computational cost normally used for the tuning of Monte Carlos, to generate search results with clusters from simulated signals and to train a deep learning neural network on these. 

Deep learning or deep neural networks are a subfield of machine learning which is inspired by the way brain works. A network consists of layers of nodes/neurones that endeavour to recognize  features in the input data.  Neural networks have proven successful in many applications, including ones in gravitational waves physics: noise studies and de-noising \cite{raz18,geo17,she17, we19}, detection of binary inspiral signals and estimation of their parameters \cite{fan19,she19,geb19,geo18,gab18,2020arXiv200704176L,2020PhRvD.102f3015S}. Recently \cite{dre19, dre20,Yamamoto:2020pus} have proposed to use deep neural networks for detecting continuous gravitational waves. See \cite{Cuoco:2020ogp} for a review.

We begun the clustering-network development in \cite{beh20} tackling the simpler clustering problem, that is to cluster candidates from loud signals. We now want to explore the clustering of weaker signals. 

The paper is organized as follows. In Section ~\ref{sec:GWresults} we recall what the general context is, for gravitational wave search results from broad surveys, and we concentrate on the broadest existing surveys, i.e. the Einstein@Home ones.  In Section \ref{sec:WeakSigN} we present the new network, capable of detecting weak-signal clusters; in Section \ref{sec:CascadeN} we discuss a composite network. Finally in Section \ref{sec:conclusion} we recap the main results, and compare and contrast the performance of our newly developed clustering schemes with existing methods. 

\section{Gravitational wave search results}
\label{sec:GWresults}

We develop and benchmark our clustering method on the results of the latest Einstein@Home all-sky search \cite{Steltner:2020hfd}. This search utilises public data from the LIGO O2 run \cite{Vallisneri15, ligoweb} spanning about 9 months, covers a signal-frequency $f$ between 20 Hz and $\approx$ 600 Hz and a spin-down $\fdot$ between  $-2.6 \times 10^{-9}$ Hz/s and $2.6 \times 10^{-10}$ Hz/s. The results stem from the combination of $64$ coherent searches, each performed on data from the two LIGO detectors over a 60 hr time-span. 

An Einstein@Home search \cite{web1}  splits the computational workload in to millions of separate work-units and distributes them among the volunteer computers.  Each work-unit  searches just under $10^{11}$ different waveforms corresponding to a 50 mHz band in frequency, the entire spin down range, and a small patch of the sky. The central Einstein@Home server receives back information only about the most significant 7500 of these waveforms. We refer to these as candidates. A candidate comprises a value of the detection statistics, and a set of template parameter: $f,\fdot,\alpha,\delta$.

The detection statistic that we use is the odds ratio ${\hat{\beta}}_{S/GLtLr}$ between the continuous signal (S) hypothesis, defined by the waveform parameters, and the noise hypothesis, given the data. The noise hypothesis is Gaussian noise (G) or a spectral line, either persistent (L) or transient (tL). For the purposes of this paper, the detection statistic can just be considered a measure of likelihood that the data contains a continuous signal with certain parameters; the interested reader can find more details about the detection statistics in \cite{Keitel:2013wga,Keitel:2015ova}.

\begin{figure}[!htb]
\center{\includegraphics[width=0.9\columnwidth]{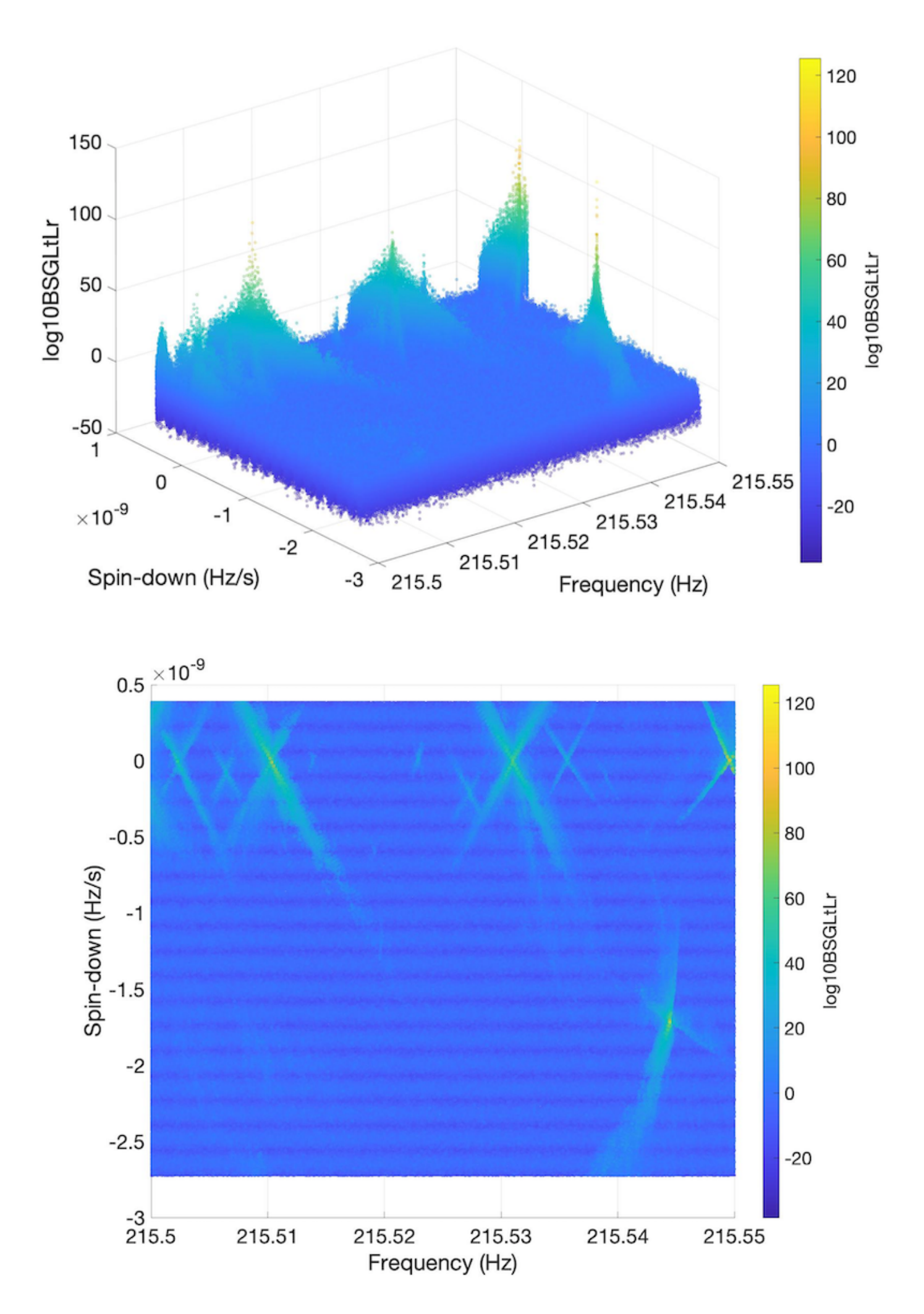}}
\caption{\label{fig1}  Results from the Einstein@Home search in the $215.50-215.55$ Hz band; the detection statistic as a function of $f$ and $\fdot$ is shown. Fake signals appear in this result-set as parameter space regions with enhanced values of the detection statistic. The lower plot displays the top-down view. 
}
\end{figure}

Figure~\ref{fig1} shows the results in the $215.50-215.55$ Hz band, as a function of the candidates' frequency and spin-down, with the detection statistic value color-coded. Several X-shaped areas are evident that present high values of the detection statistic: these are due to loud fake signals added to the data before the search. Other than for these regions, the input data to our network can be broadly thought of, as images like this.

\section{The weak-signal network}
\label{sec:WeakSigN}
\subsection{Target signals}
\label{sec:WeakSigNTargets}
In \cite{beh20} we developed a network that could cluster search results from loud signals, giving rise to structures like the ones shown in Figure~\ref{fig1}. We refer to this network as the loud-signal-network (LoudSigN). The detection efficiency of this network drops to less than 17\% for signals with detection statistic ${\hat{\beta}}_{S/GLtLr}\sim 20$, and this is inadequate for the level of sensitivity of the latest searches: in  \cite{Steltner:2020hfd} nearly all candidates that are followed-up have ${\hat{\beta}}_{S/GLtLr}$ values lower than 20.

The input data for the large-signal-network is ``down-sampled" : the resolution of the original $f-\fdot$ images is reduced by averaging over a certain number of pixels, $12\times12$ for the O2 data. This is a necessary step because it blurs away the structure of local ``peaks and valleys" making it easier for a network to clump everything together as a single cluster. On the other hand this step also decreases the contrast of the signal peak, and if the signal is weak enough, this step blurs it away completely.

\begin{figure}[!htb]
\center{\includegraphics[width=1\columnwidth]{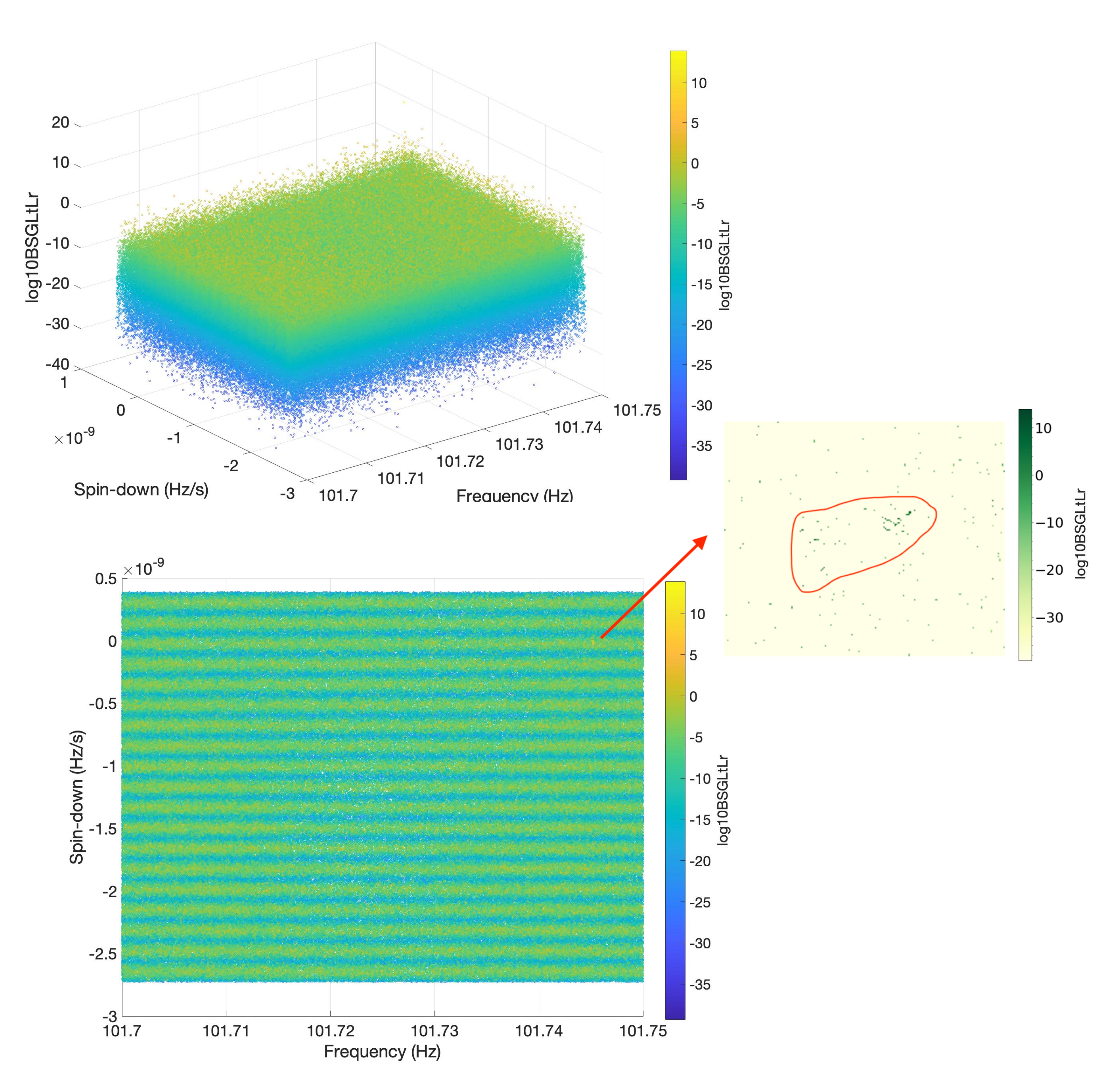}}
\caption{\label{fig1w}  Results from the Einstein@Home search in the $101.70-101.75$ Hz band; the detection statistic as a function of $f$ and $\fdot$ is shown. These plots are like the ones shown in Figure~\ref{fig1}, apart for the fact that the fake signal added to the data is not as loud as any of the ones of Figure~\ref{fig1}. In fact, this signal is well below the detectability level of the LoudSigNet, and it is barely visible by eye in the results output. By comparing this figure with Figure~\ref{fig1}, we show how different the target clusters from loud and weak signals are, which illustrates why we need a specific network for the weak signals. The zoomed-in panel shows the region where the signal cluster is, with the red lines delimiting the ground-truth cluster region. Note that the color-coding for the zoomed-in panel is different than for the main figure.}
\end{figure}

We design a network aimed at these weaker signals. We define these signals as those that would not be visible in the input data prepared for the loud-signal-network, but that are visible or barely visible in the original O2 data. We refer to this new network as the weak-signal-network (WeakSigN). Figure \ref{fig1w} shows an example of a weak target signal. 

\subsection{Input data and ground truth}
\label{sec:InputDataGroundTruth}
The input to the network are images with $512\times512$ pixels, that can be handled by the high-end 32 GB graphical processing unit (GPU). In this result-set they correspond to slices of 1.7 mHz $\times$ 1.7 $\times$ 10$^{-10}$ Hz/s. 

As explained in Section \ref{sec:GWresults}, the Einstein@Home results are ``top-lists" of the 7500 highest-ranking candidates from each portion of parameter space explored. Since each portion explores of order $10^{10}$ waveforms, this means that no results are returned for most of the parameter space. In our $f,\fdot$ network-input images we typically find over 99\% of empty pixels. When a pixel in the input image is empty, we assign to it a random score that is $\lesssim$ the lowest value of the detection statistic in the image. In Figure \ref{fig1w} such value is  $\sim -39.3$ and there are $132051137$ empty pixels out of $132524538$. The empty pixels can be seen in the zoomed-in panel, they are the yellow area. 
It is possible that more than one candidate is found in the same $f,\fdot$ pixel, corresponding to different sky positions. In such cases, as done in \cite{beh20}, the candidate with the highest detection statistic value across the sky, is picked.

We use a supervised network, which requires with each training input the corresponding output, called ground truth. To generate the training set, we add weak fake signals, like the ones that we want to cluster, to the raw LIGO data. The signals' frequency and spindown are within the search band of the original search, between $20-600$ Hz, and $-2.6 \times 10^{-9} - 2.6 \times 10^{-10}$ Hz/s, respectively. The right ascension and the cosine of the declination of the source, the cosine of its inclination angle and the polarization angle \cite{Jaranowski:1998qm} are distributed uniformly in $[0, 2\pi) , [-1, 1), [-1,1)$ and $[ -\pi /4 , \pi /4)$, respectively. We run a search like the Einstein@Home search on this data and from the results we produce the corresponding network-input image. We construct the ground truth by selecting by eye the clustered pixels that contain traces of the signal. Even though these are barely visible, knowing ``where to look", i.e. knowing the signal parameters, makes this possible. We use an image editing tool (Pixelmator) on a tablet computer equipped with a touchscreen. We load each training image in the editing tool and mark the cluster region -- an example is given in the zoomed-in panel of Figure \ref{fig1w}. This information is converted in the ground truth matrix ${\cal{T}}_{ij}^\alpha$, with $(i,j)$ labelling the $(f,\fdot)$ pixels of the image and $\alpha=1\cdots N_{\textrm{cl}}$ labelling the signal clusters of that image. ${\cal{T}}_{ij}^\alpha=1$ if that pixel is part of the $\alpha$ cluster, and zero otherwise.

\subsection{The network}

The clustering method is addressed by using instance segmentation networks. The general network architecture, the overall data structure, and preparation of the input data set for the network is similar to  \cite{beh20}. Here we omit the details already described in  \cite{beh20} and instead concentrate on the key novelties and results.

\begin{figure}[!htb]
\center{\includegraphics[width=0.4\columnwidth]{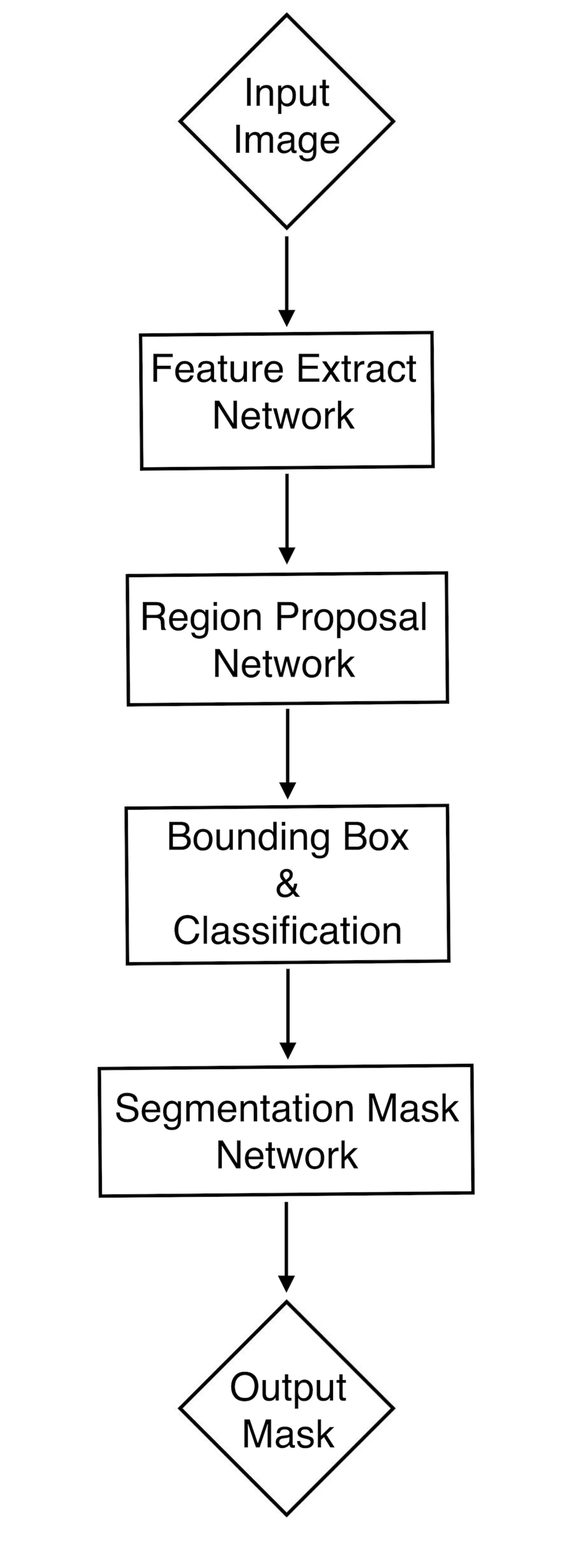}}
\caption{\label{fig2} A schematic diagram of the Mask R-CNN architecture used for our clustering networks.}
\end{figure}

The network scans the image and finds the regions that most likely include the cluster, then classifies them and generates definitive boundaries. The output is a pixel mask that determines the boundaries and a score that identifies how likely a pixel is part of a signal cluster. The score threshold that decides whether a pixel is or not part of a cluster is set at 0.5. The network structure  is summarized in Figure~\ref{fig2}.  

In order to train the network, we need to set model weights and batch size.  Model-weights are the set of coefficients and biases that define how each node in the network transforms the data. We set the starting model-weights to be the weights of the trained network in \cite{beh20}. The training input data is divided into several batches with a size defined by the batch size parameter. The network works through all samples in each batch  before updating the network weights. The batch size is mainly limited by the size of the GPU memory.  In  \cite{beh20}, a large batch size of 15 made a big improvement in the network performance because the network started with generic model weights that were not specific to our problem. Here we start with model-weights from  \cite{beh20} and this helps to significantly lower the batch size without loosing in performance and at the same time frees up memory to use a bigger size sub-image. Having bigger sub-images gives a better picture of the clustered structures over the background noise that embeds them. For this network we use a batch size of 2 with sub-images of $512\times512$ while in \cite{beh20} the batch size was 15 with sub-images of size $256\times 256$.   

Similar to  \cite{beh20}, the training is performed in  3 steps which proved to enhance performance. Since we begin with more reliable weights, the first two training steps can be made shorter compared with  \cite{beh20}.  The steps are as follows: 1) only the first layer of the three last levels in Figure~\ref{fig2} is trained, for 60 epochs rather than 100 epochs  2) using the weights from the previous step, the first three levels are trained for 120 epochs rather than 300 epochs 3) the complete network is trained with the weights from the second step for 1000 epochs. The best performance of the network is achieved with these hyper-parameters: learning rate=0.001, weight decay=0.00001, learning momentum=0.9. The training process takes about 40 hours to complete, which is much less than the time that it typically takes to tune a deterministic clustering algorithm.

More details on the choice of the network parameters is provided in Appendix \ref{app:networkParams}.

\subsection{Results}
\label{sec:result}
The network, with the configuration explained in the above section, is trained on 515 sub-images and validated on 218 sub-images. The  network is tested on 347 sub-images, totally independent of both the training and the validation set. Each image contains a weak signal cluster, as described in Section \ref{sec:WeakSigNTargets}. The underlying signal parameter distributions reflect the target signal population and are those described in Section \ref{sec:InputDataGroundTruth}.

\begin{figure}[!htb]
\center{\includegraphics[width=0.9\columnwidth]{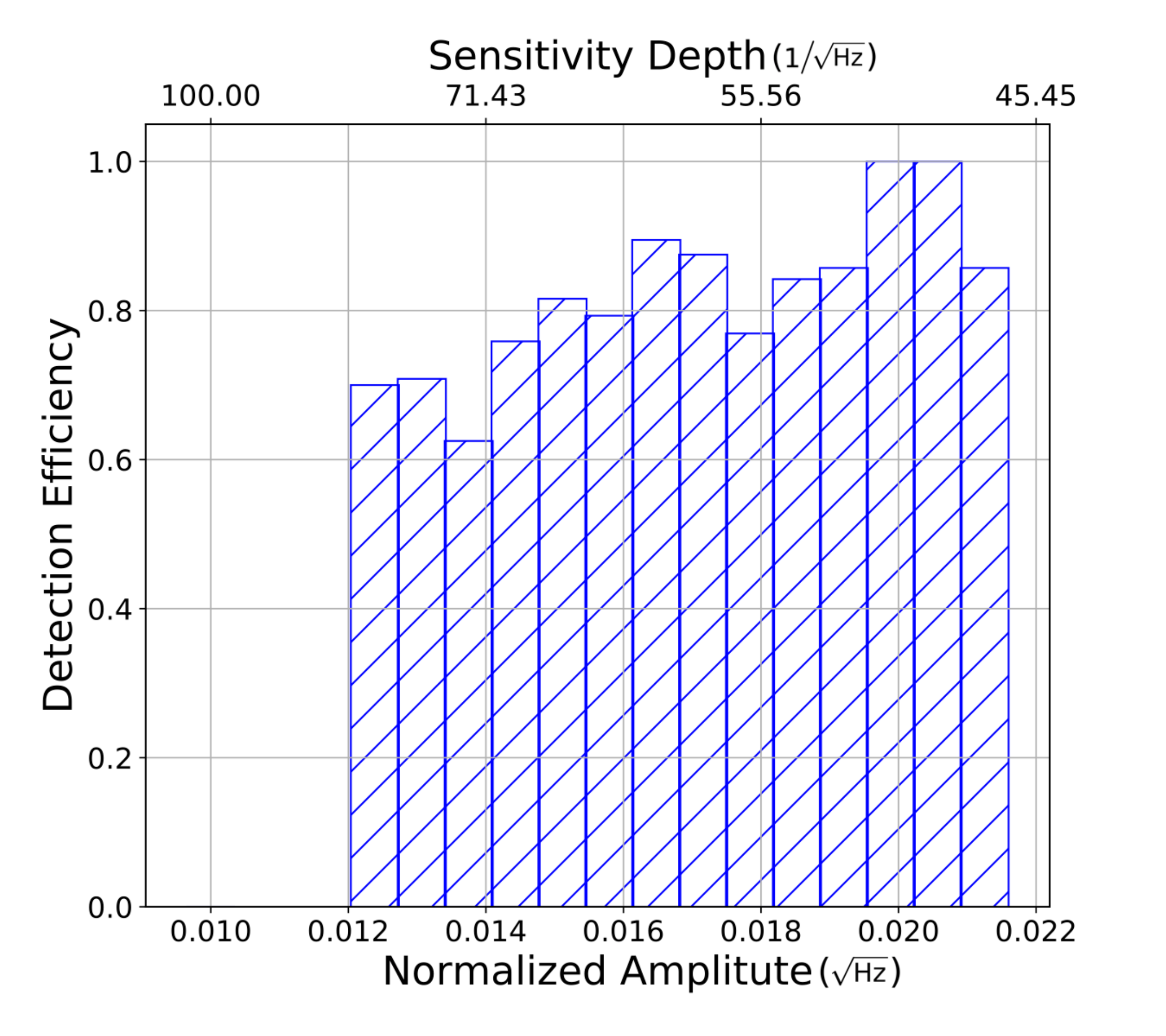}}
\caption{\label{fig3} Detection efficiency of the WeakSigN as a function of fake signal normalized amplitude.}
\end{figure}

The network correctly identifies 276 of such clusters, corresponding to 80\%  detection efficiency, and with a statistical error of $\pm$ 2\%. The detection efficiency as a function of the normalized signal amplitude is shown in Figure~\ref{fig3}. We define the normalized amplitude for a signal at frequency $f$ with intrinsic amplitude $h_0$ the ratio $h_0/\sqrt{S_h(f)}$, where $S_h(f)$ is the power spectra density at that frequency. We note that the normalized amplitude is the inverse of the sensitivity depth that a search should reach in order to detect this population of signals, and has thus units of [$\sqrt {\textrm{Hz}}$].

\begin{figure}[!htb]
\center{\includegraphics[width=0.9\columnwidth]{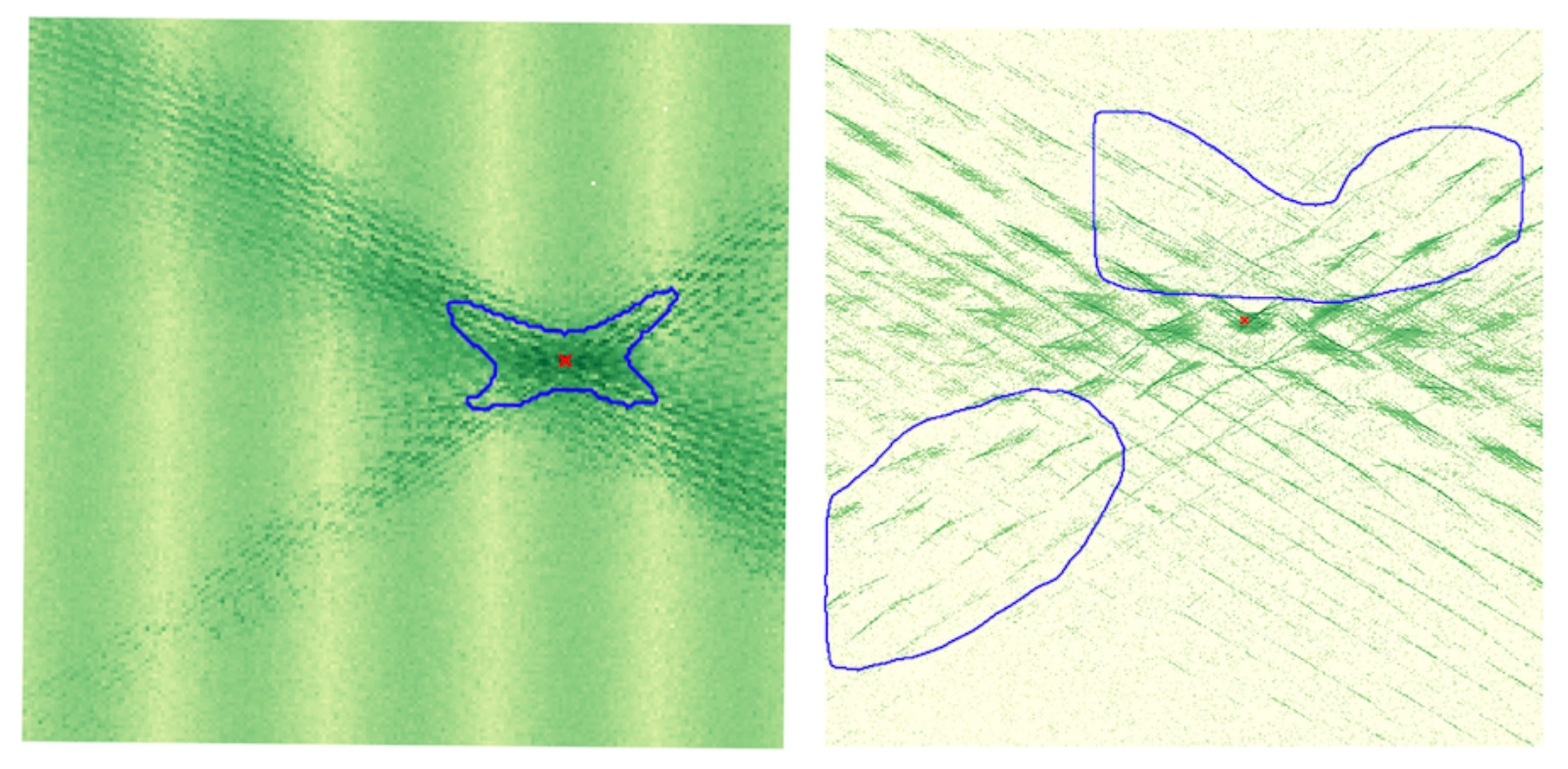}}
\caption{\label{figLS} Performance of WeakSigN (right image) and LoudSigN (left image) on one loud signal cluster at frequency of 459.035 Hz. The blue lines show the detection boundaries of the clusters found by the network and the red cross shows the injection. Left image shows the signal cluster in low-resolution image and the right image shows this cluster in high-resolution image. The LoudSigN observed one region contained the signal cluster while the WeakSigN detected 2 clusters for that.}
\end{figure}

The WeakSigN does not perform as well on loud signals. Clusters from loud signals typically spread over more than a single sub-image and present structures. The network picks up parts of these structures as being small clusters, which generates a higher rate of false alarms around loud signals and disturbances. In other words, the network misses the general picture. We show an example of this in Figure~\ref{figLS}, where the network does not even return a cluster that contains the signal parameters.

\section{The cascade-network}
\label{sec:CascadeN}

We investigate the combination of the LoudSigN and the WeakSigN to efficiently detect signal clusters over a broad range of signal amplitudes. We re-train the LoudSigN of \cite{beh20} on the O2 data and we use it on the O2 data. We record the clusters identified by this network and set the corresponding pixels to ``empty" in the original image. The resulting image is fed to the WeakSigN. We record the clusters identified by the WeakSigN.

\begin{figure}[!htb]
\center{\includegraphics[width=0.9\columnwidth]{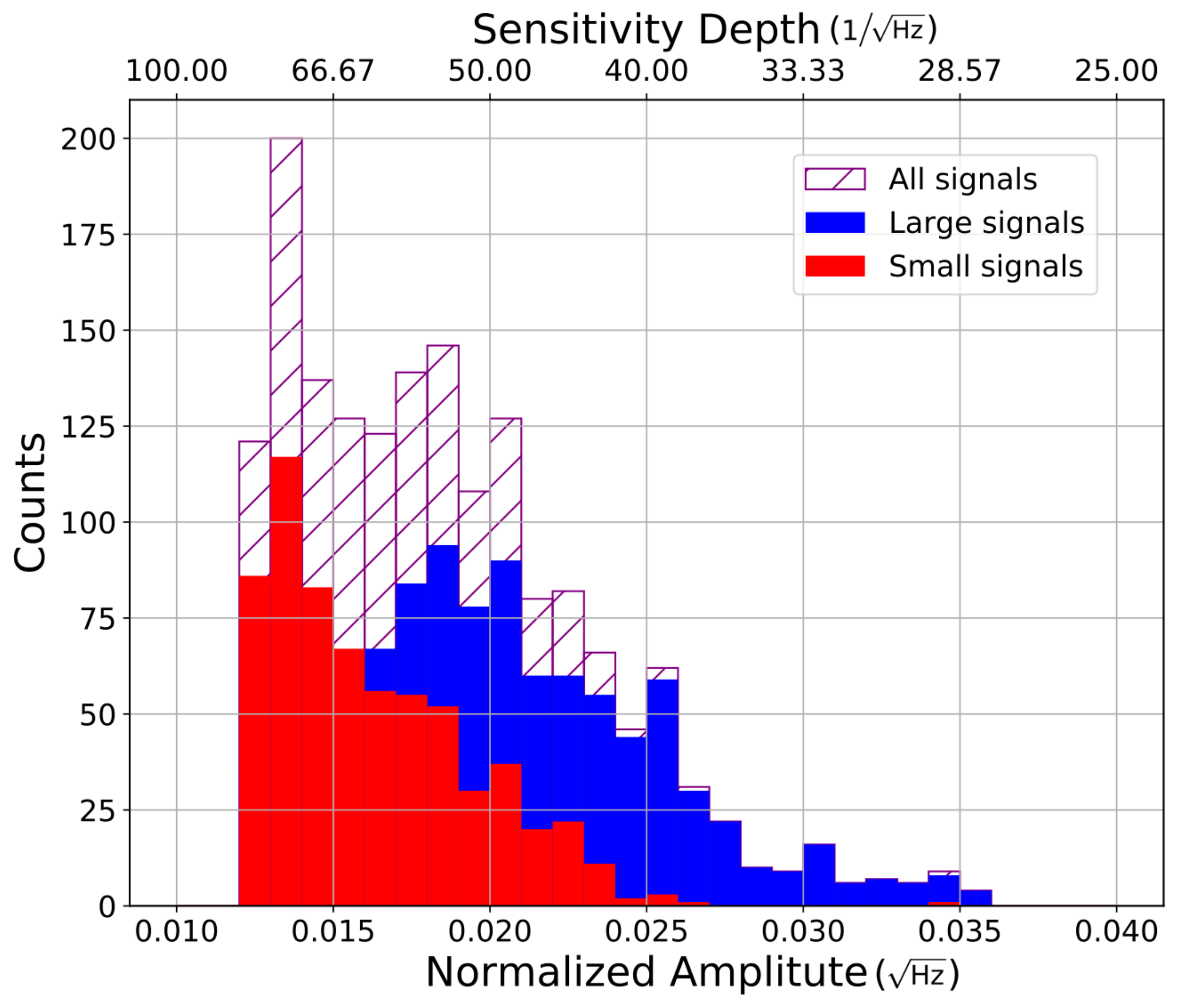}}
\caption{\label{fig4} Distribution of the normalized gravitational wave amplitudes of the fake signals used to characterise the performance of the cascade-network. The different colours shows how the signal appeared in the search output, i.e. whether it gave rise to a loud cluster or a weak cluster.}
\end{figure}

\begin{figure}[hbt]
\includegraphics[width=0.9\columnwidth]{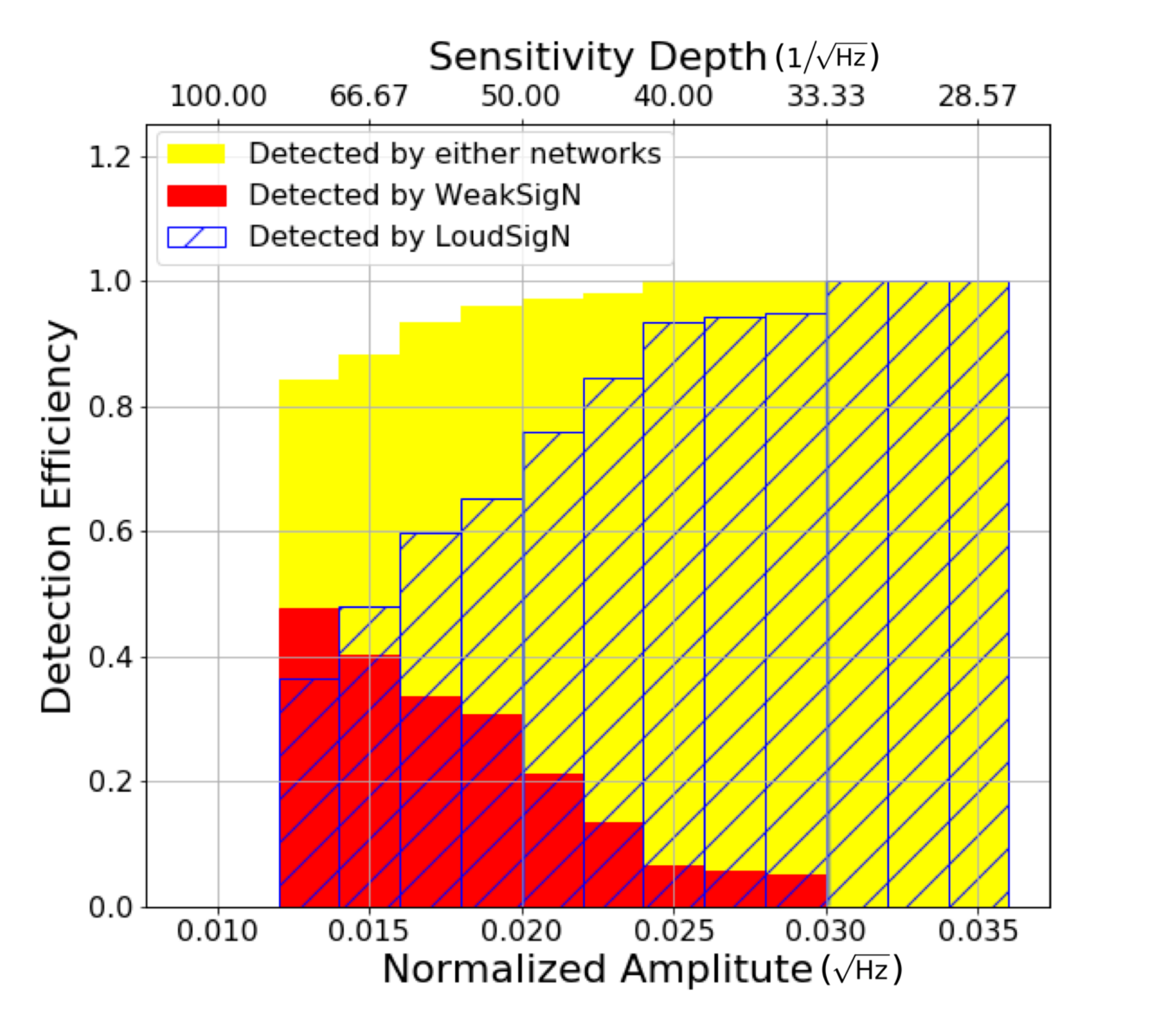}
\caption{\label{fig5} Top: Detection efficiency of the cascade method network. The distribution of each network is shown in different colours. }
\end{figure}

We characterize the performance of the cascade-network on 1684 fake signals in the frequency range 40-579 Hz and having a wide range of signal strengths, as shown in Figure~\ref{fig4}. We consider a signal detected when the cluster contains at least a candidate from the signal. It may happen that the first network identifies part of a signal cluster, and the second network picks-up weaker candidates. In this case the same injection is associated with a cluster in each network.
Figure~\ref{fig5} shows the efficiency in detecting the clusters associated to the injected signal, as a function of the signal strength. 

We find that all clusters with normalized signal amplitude of $\gtrsim$ 1/40 $\sqrt {\textrm{Hz}}$ are detected and that the network has a detection efficiency of more than 95\% up to  signal strengths of $\approx$ 1/54 $\sqrt {\textrm{Hz}}$. For weaker signals the new network significantly contributes to the detection efficiency and at $\approx$ 1/80 $\sqrt {\textrm{Hz}}$ accounts for more than half of the detected signals.

\begin{figure}[hbt]
\includegraphics[width=.7\columnwidth]{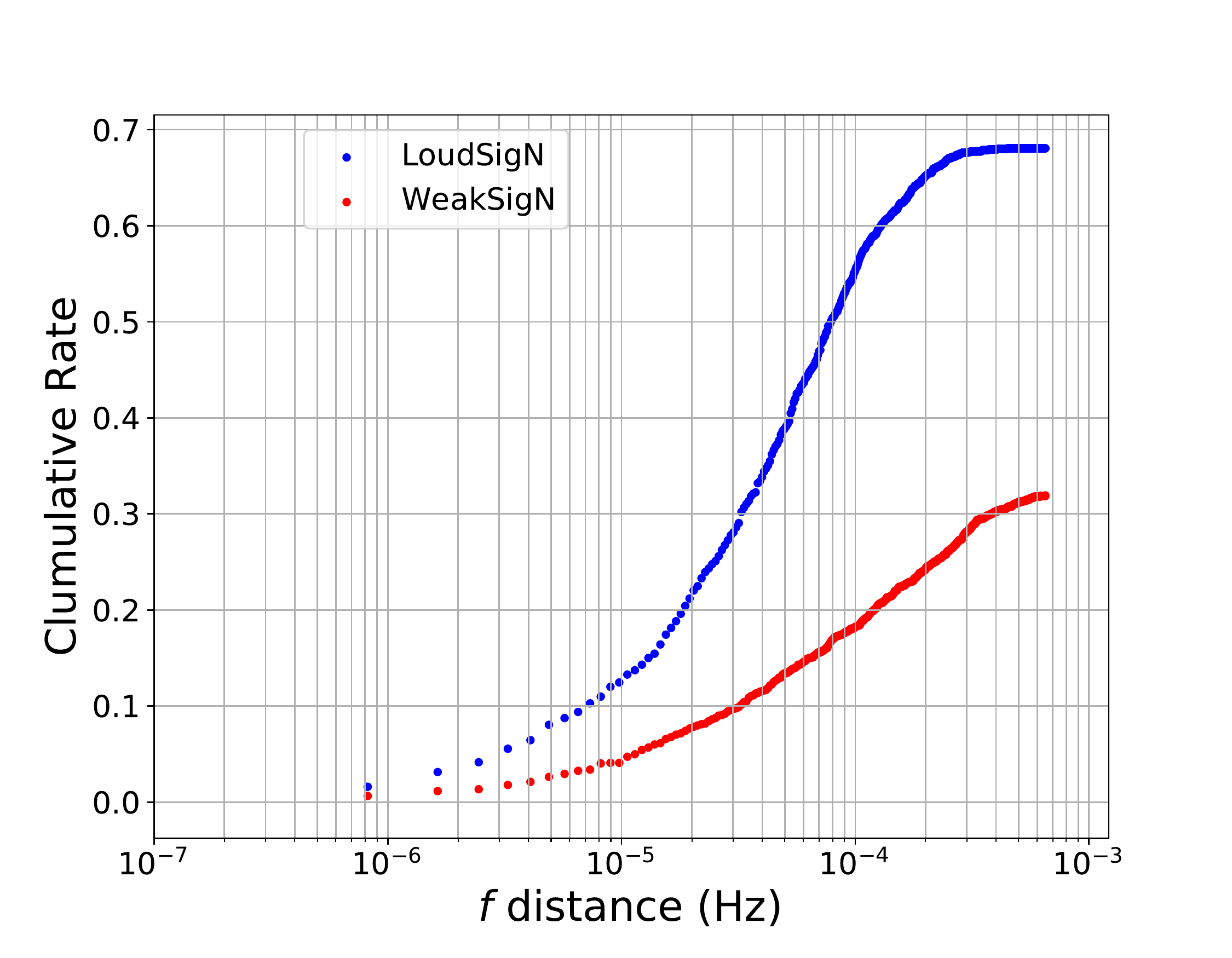}
\includegraphics[width=.7\columnwidth]{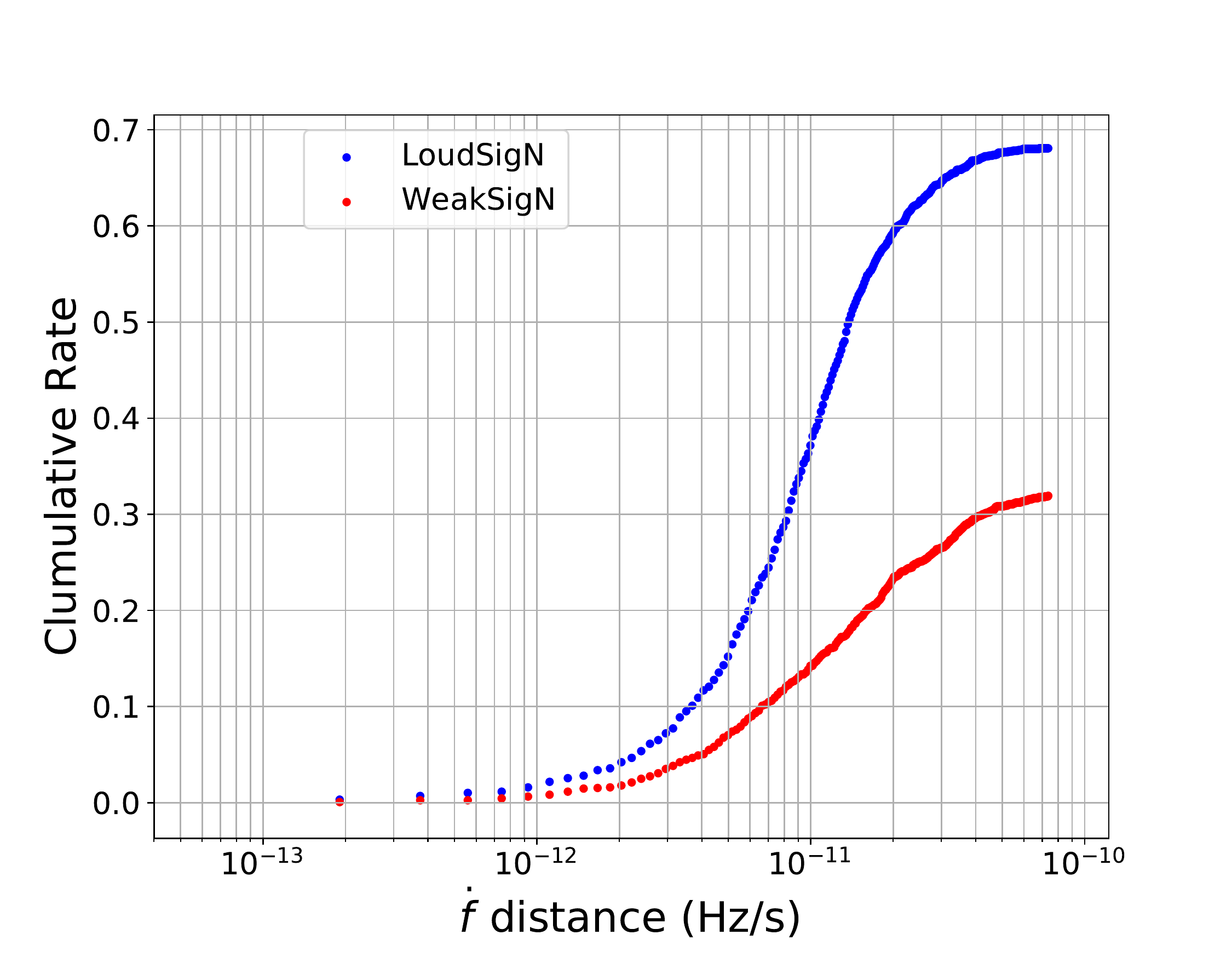}
\caption{\label{fig6} Cumulative distribution of the  distance between the signal parameters and the parameters of the most significant cluster candidate (the candidate with the highest detection statistics value) recovered by the network. The top plot shows the distance in frequency; the bottom plot the distance in spin-down (right).}
\end{figure}

We evaluate the uncertainty in signal parameters for each signal identified by the network. 
We use the results from the fake signals studies as follows: We find the candidate corresponding to the highest value of the detection statistic in each signal cluster, and calculate the distance to the actual signal parameters. If more than a cluster exists associated with the same fake signal, for the purposes of evaluating this distance, we consider the cluster that is closest to the signal. 
Figure~\ref{fig6} shows the cumulative rate of candidates within a given $f,\fdot$ distance of the actual signal parameters.  
Our results indicate an uncertainty region of $4.5\times10^{-4}$ Hz and $6.8\times10^{-11}$ Hz/s for cluster from the LoudSigN and of $6.5\times10^{-4}$ Hz and $7.4\times10^{-11}$ Hz/s for clusters identified by the WeakSigN. 

We evaluate the false alarm rate of the cascade-network by running it on the Einstein@Home O2 result-set \cite{Steltner:2020hfd}. We randomly pick 1774 50-mHz frequency bands over the entire search range and apply the cascade-network to the results from these bands. The false alarm rate is dominated by the  LoudSigN and the WeakSigN shows a negligible false alarm rate.

\begin{figure}[!htb]
\center{\includegraphics[width=1\columnwidth]{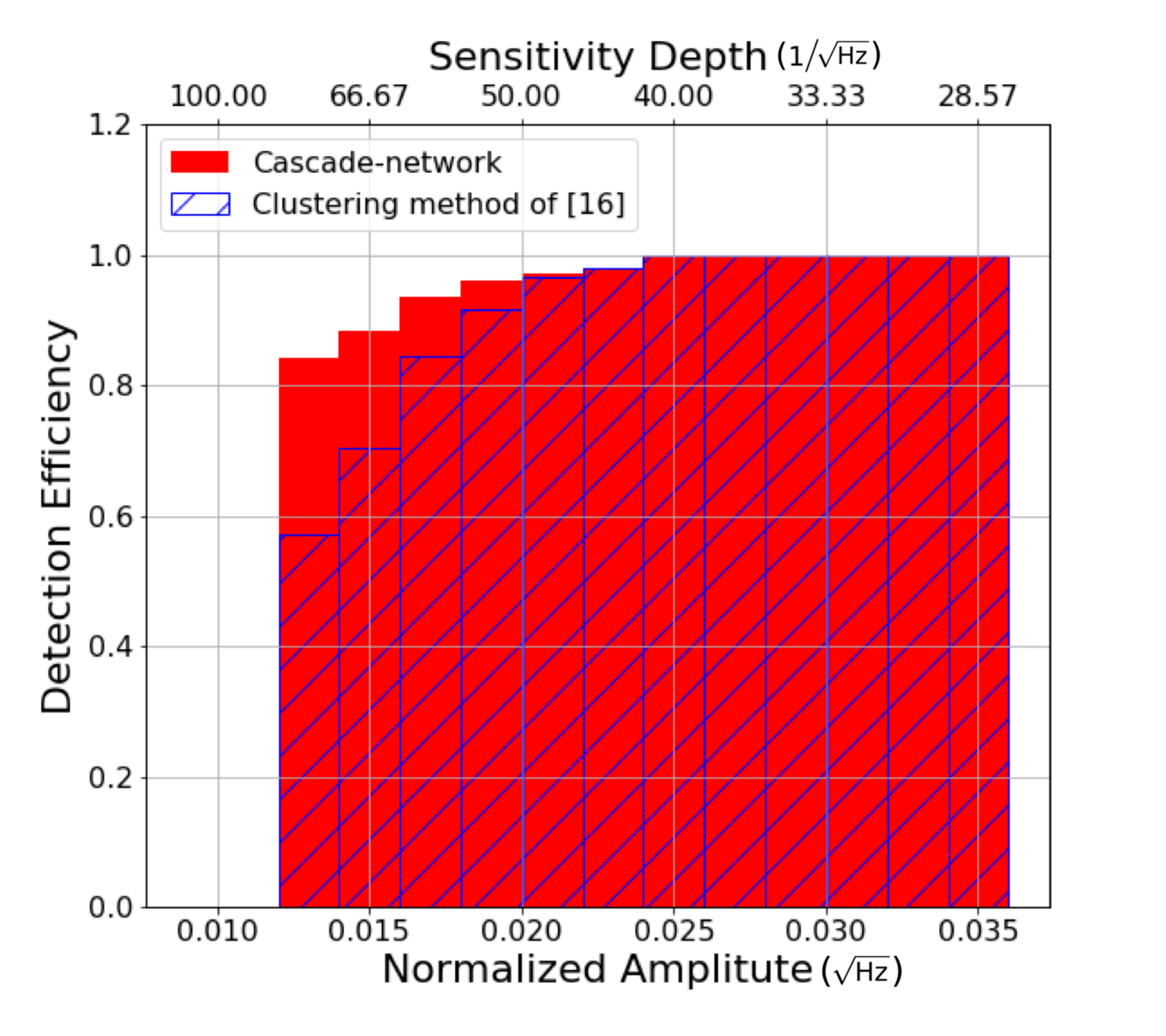}}
\caption{\label{fig8} Comparing the cascade-network with the deterministic clustering method used in the analysis of the Einstein@Home search results \cite{Steltner:2020hfd}. Our network shows a higher detection efficiency for lower amplitude signals.}
\end{figure}

Overall our cascade network generates about half the false alarms than the clustering method employed in the latest Einstein@Home all-sky search on the same bands \cite{Steltner:2020hfd}. 

\begin{figure}[!htb]
\center{\includegraphics[width=0.9\columnwidth]{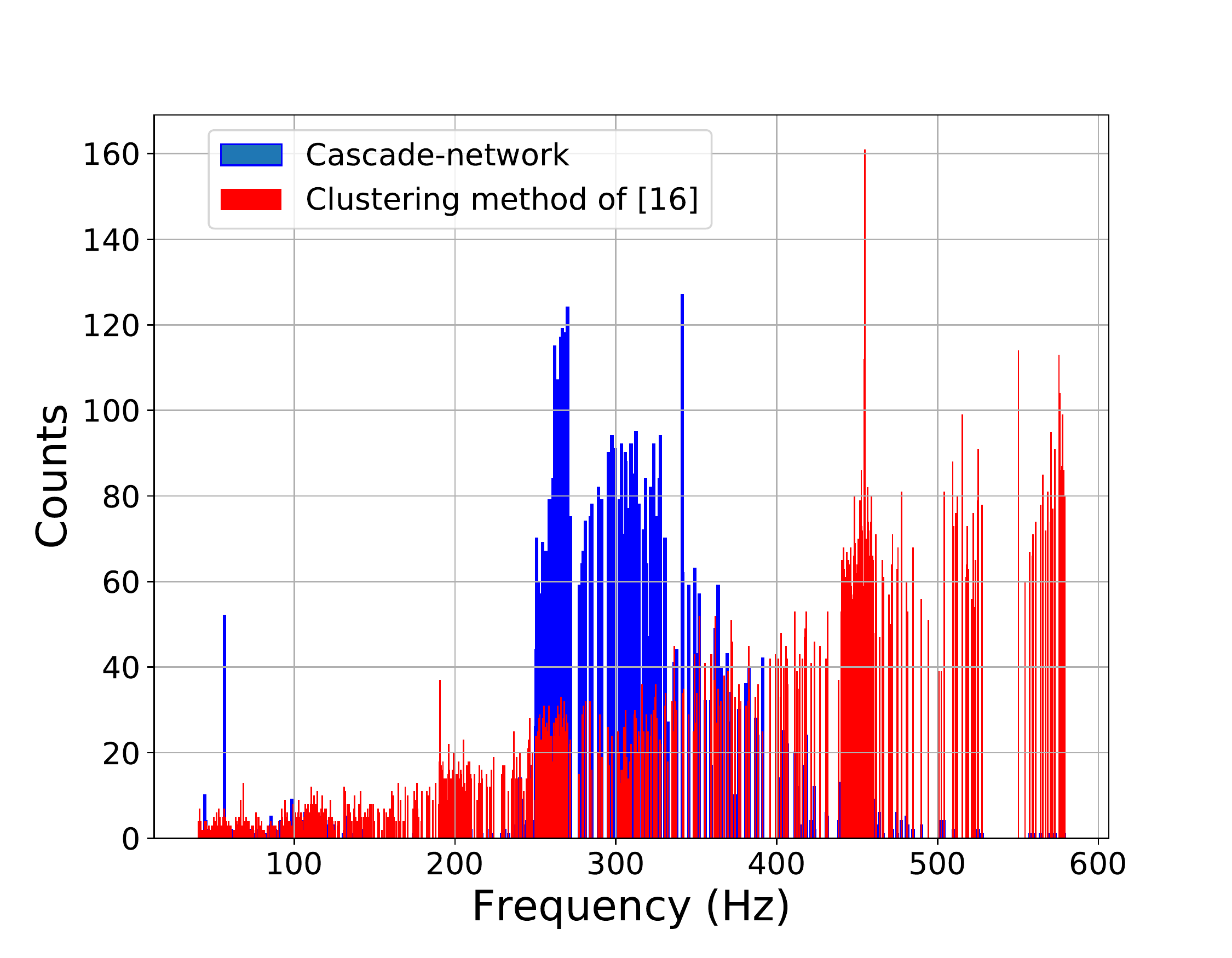}}
\caption{\label{fig7} False alarm rate versus signal frequency for our network and for the clustering method used in \cite{Steltner:2020hfd}. }
\end{figure}

As shown in Figure~\ref{fig7}, the false alarm rate of the cascade network is not constant in frequency. The higher false alarm in the middle frequency range stems from the stripe-features visible in Figures~\ref{fig1} and \ref{fig1w}, that display lower detection statistic values in the result-set appearing periodically at constant $\fdot$. The origin of these stripes is well understood and comes from the ``recalculation" of the detection statistic of the candidates of the top list (see Appendix \ref{app:stripes}). Since the number of templates searched in a 50 mHz band increases with frequency, these stripes are more pronounced with increasing frequencies. The WeakSigN is not sensitive to them because it works on small sub-images such that the stripes just produce a constant background level that is irrelevant. The LoudSigN, instead, works on larger sub-images and can resolve the stripes. The appearance of the stripes and of the signal clusters changes with frequency and the training-sets must include enough ``models" of these different types of behaviour, in order to properly distinguish signal clusters from other features. The 250-350 Hz region is a transition region where the images become very smooth, with a much higher pixel density than at lower frequency. The higher false alarm that we see stems from our training-set on the O2 data being somewhat coarse in this mid frequency range, and can be remedied by increasing it in this region.

\section{Conclusion}
\label{sec:conclusion}

In \cite{beh20} we presented the first deep learning network trained to identify signal clusters in the output of very broad searches for continuous wave signals. That network was aimed at large signals. In this paper we build on this and design a second network that is capable of identifying clusters from faint signals. We train and test the network on the most recent results from an all-sky Einstein@Home search of LIGO O2 data \cite{Steltner:2020hfd}. We show that a cascade-architecture of these two networks can identify clusters of continuous gravitational wave signal candidates over a broad range of signal strengths.

The cascade-network shows an excellent performance, with a detection efficiency of 92\%  above normalised signal amplitudes of 1/60 $\sqrt {\textrm{Hz}}$. This exceeds the performance of the deterministic clustering method used in \cite{Steltner:2020hfd} which has a detection efficiency of 83\% in that amplitude range.  For weak signals -- at normalised amplitudes of 1/80 $\sqrt {\textrm{Hz}}$ -- our cascade-network maintains a detection efficiency of $\gtrsim 80\%$ whereas the deterministic clustering performance drops to less than 50\%. Figure~\ref{fig8} summarises the performance comparison. 

Over a broad frequency range the overall false alarm rate of our cascade-network is half that of \cite{Steltner:2020hfd}, albeit with peaks up to four times larger than \cite{Steltner:2020hfd} between 250 Hz and 350 Hz. These false alarms are due to the stripe-features in the results-set. We used a sparse distribution of training-set signals at a few sample frequencies, and the range between 250 Hz and 350 Hz was not very well covered. We are confident that a training set for the LoudSigN with adequate coverage of this range would train the network to distinguish between signals and stripes and hence reduce the false alarm rate in this band. One could also imagine pre-processing the images to suppress the stripes. However, since what really matters is the overall false alarm rate, because that determines the computational cost of the next stage, the higher false alarm rate in the mid frequency range does not invalidate the method.

The uncertainty in $f$ and $\fdot$ of our network is competitive, being 1.68 and 1.25 times smaller than the uncertainties of the clustering method used in \cite{Steltner:2020hfd}.

As explained earlier on, each template is identified by a sky location as well as a $f$ and $\fdot$ value. The clustering described in this paper applies to $f$ and $\fdot$ and is immediately relevant to continuous wave searches such as \cite{Papa:2020vfz, Ming:2019xse, Zhu:2016ghk}, where the sky position is informed by X-ray observations with excellent accuracy and only a single sky position is searched. For an all-sky search the question naturally arises of how well the network identifies the sky position of a signal, even without explicit clustering in the sky. The uncertainty in the sky by LoudSigN is $\approx$ half that of the deterministic clustering method; but the sky-uncertainty of the WeakSigN is much larger than that of the deterministic clustering method, in occasions yielding seemingly unrelated sky positions. In this respect the sky-localisation performance of the cascade network falls short of optimal. A recent study demonstrates that a deterministic clustering based on the density of candidates in 4D cells ($f, \fdot$, [sky position]) works, even on weak signals \cite{Steltner:2020hfd}. This indicates that a deep-learning clustering able to confidently identify weak signals should use the density of candidates rather than the detection statistic, and should consider all dimensions at once. The latter point likely involves a significant change in the construction of the training set and perhaps in the network architecture. This will be our next focus and we believe that it will yield a deep-learning clustering for the results of very broad all-sky continuous wave searches.

\section{Acknowledgments}
We thank Benjamin Steltner for providing the performance benchmark data for the clustering procedure used in \cite{Steltner:2020hfd}. The computing work for this project was carried out on the GPUs of the Atlas cluster of the Observational Relativity and Cosmology division of the MPI for Gravitational Physics, Hannover \cite{obsrelDiv}. We thank Bruce Allen for supporting this project by granting us access to those systems. 

This research has made use of search results from our previous analyses that utilize LIGO data from the LIGO Open Science Center (\url{https://losc.ligo.org}), a service of LIGO Laboratory, the LIGO Scientific Collaboration and the Virgo Collaboration.  LIGO is funded by the U.S. National Science Foundation. Virgo is funded by the French Centre National de Recherche Scientifique (CNRS), the Italian Istituto Nazionale della Fisica Nucleare (INFN) and the Dutch Nikhef, with contributions by Polish and Hungarian institutes.

\appendix

\section{Origin of the stripe-feature in the Einstein@Home result set}
\label{app:stripes}

For the reader who is familiar with semi-coherent continuous wave searches and interested in the technical details, in this appendix we explain in more detail the origin of these stripes. We also refer to Section 3.2 of \cite{Steltner:2020hfd}. The recalculation of the detection statistic consists in computing the detection statistic in each coherent search at the $\fdot$ template value of the candidate, so that the resulting average is not any more an approximation of the average detection statistic at such grid point. The recalculation is performed because statistically it increases the detection statistic value of a signal with respect to noise. This step introduces a systematic effect in the noise, depending on whether the original top-list candidate had a single-segment coarse-grid $\fdot$ value close to the final fine-grid template value or not. For those candidates with coarse- and fine- grid $\fdot$ templates close to each other, the recalculation has no effect and leaves the detection statistic value unchanged, and hence high. For those candidates for which the coarse- and fine- grid points are ``further away" the approximated and exact detection statistic differ more and in noise the recalculated value is lower. This is what generates the modulation in recomputed detection statistic values as a function of $\fdot$, i.e. the stripes.

\section{Choice of network parameters}
\label{app:networkParams}
We chose a MaskRCNN network because this type of architecture has proven very effective in instance segmentation problems, as our cluster identification. We have illustrated the general network architecture in \cite{beh20}.

Every deep-learning network has a number of parameters that must be set. Optimizing performance and computing cost is a non trivial endevour, especially for complex architectures such as the R-CNN deep-learning network. Producers of GPUs even make hardware design choices based on the deep learning network performance for certain problems -- see for instance \cite{aws}.

A full-scale optimisation study is well beyond the scope of this work. The scope of this paper is to present a viable and reasonable R-CNN network set-up and demonstrate that a deep learning  approach lends itself to avoiding the cumbersome tuning of deterministic algorithms. With ``viable and reasonable" here we mean that the detection performance is comparable/better with that of deterministic approaches and that using the network on a large data set (i.e. from a real Einstein@Home search) would be possible within a time-scale comparable/shorter than that of deterministic approaches.

In this appendix we show that the operating points that we have chosen are reasonable, and even close to optimal within the constraints that we operate under.

The size of the sub-image is probably the most important parameter for our problem (in this paragraph when we say ``image" we mean ``sub-image"). The performance of the clustering network increases when the image is larger, because larger images are more likely to contain edges, which is what the network firstly detects. As we explain in Section IIIA, resolution is also important, as a higher resolution mean higher contrast for the features that the network identifies. The limit to image size is set by the GPU that we have: a 512$\times$512 image is manageable by our high-end 32GB GPU, but an image 1024 $\times$ 1024 is not. This sets a hard limit to the image size. A 256 $\times$ 256 image is not manageable by a 6GB GPU whereas a 64$\times$64 image, is.

Sub-image size in principle impacts the training time, but within the range of viable sub-image sizes not so much, because the operations are very effectively parallelised on the GPU: for an image with x 16 more pixels (from 128$\times$128 to 512$\times$512) the training time becomes only 1.6 times longer.

The training set size impacts the training time, which scales linearly with the number of samples (sub-images). Every sub-image may contain more than a signal, and the network learns with every new signal. The trade-off on the size of the training set was made 
based on the significant effort required to generate the ground truth for every signal. This is done by hand as described in Section \ref{sec:InputDataGroundTruth}, for every signal, and takes, say, 4 minutes. This means that establishing the ground truth for 100 signals takes nearly 7 
hours of concentrated work. Devoting about a week to this is still competitive to the tuning of clustering algorithms and provides several hundred signals to train the network on. 

The detection efficiency also depends on the size of the training set. The training set should be broad and complete, and more training samples may help increase the detection efficiency of the network. The preparation of the training samples is however a time consuming process. So, there is a trade-off between the time it takes to prepare the training samples and the improvement that these extra samples produce on the detection efficiency. We prepared a set of training samples and tested the network with it. Then added extra training samples and re-evaluated the performance of the network. We repeated the process until we saw no significant further improvement. For example,128 images in the training set results in 71\% detection efficiency, 257 images  results in 78\% detection efficiency, and 515 images, results in 80\% detection efficiency.

The hyper-parameters that we set are the learning rate, learning momentum and weight decay. These parameters mostly affect how the network learns from the training samples, and only to a lesser degree the training time. The larger the value of the learning rate and learning momentum are, the faster the training is. But very fast training can cause overfitting and reduce the performance, so these values needs to be tuned. The weight decay parameter does not directly affect the training time and is used to increase performance when overfitting happens.

The learning rate is set to 0.001. A value 5 times higher results in very fast and unstable training and the network had the overfitting problem only after 30 minutes of training -- we recall that the training time for our network is 40 hours. A 5 times smaller value increases the training time by 15\% and decreases the detection efficiency slightly, by about 3\%. A gain of 6 minutes in training time is not significant, whereas a 3\% gain in detection efficiency is comparable to what was achieved by increasing the number of training samples from 257 to 515.

The learning momentum for our network is set to 0.9. Similar to the learning rate, higher values made the training very fast but unstable, with the network very quickly showing signs of overfitting.  For example for a learning rate of 0.99 the network becomes unstable very quickly, $\approx$ 3 minutes after the training begins. Smaller value increase the training time slightly but cause a lower detection efficiency. For example a learning momentum of 0.7 results in 72\% detection and the training time increases by $\approx$ 7\%. 

The weight decay is set to 1e-4. A 5 times smaller value doesn't significantly change the detection efficiency. A 5 times larger value yields a lower detection efficiency: 72\% rather than 80\%. The training time stays almost the same.

\bibliography{Bibliography}

\end{document}